\documentclass[twocolumn]{aastex701}
\usepackage{graphicx}
\usepackage{txfonts}
\usepackage{amssymb}
\usepackage{siunitx}
\usepackage{nicefrac}

\newcommand{\Halpha}{{\ensuremath{\text{H}{\alpha}}}}
\newcommand{\Hbeta}{{\ensuremath{\text{H}{\beta}}}}

\newcommand{\oiii}{{\text{[O\,{\sc iii}]}}}

\newcommand{\panbhs}{\text{PAN-BH*-1}}
\newcommand{\cliff}{\text{\it The Cliff}}
\newcommand{\mombhs}{\text{MoM-BH*}}
\newcommand{\caperslrd}{\text{CAPERS-LRDz9}}

\begin{document}

\title{A Black Hole Star at Cosmic Noon: Extreme Balmer break, photospheric continuum, and broad absorption by thick winds in a Little~Red~Dot at~z~=~1.7}

\author[orcid=0000-0001-5586-6950]{Alberto~Torralba}
\affiliation{Institute of Science and Technology Austria (ISTA), Am Campus 1, 3400 Klosterneuburg, Austria}
\email[show]{alberto.torralba@ista.ac.at}  

\author[orcid=0000-0003-2871-127X]{Jorryt~Matthee}
\affiliation{Institute of Science and Technology Austria (ISTA), Am Campus 1, 3400 Klosterneuburg, Austria}
\email{jorryt.matthee@ist.ac.at}

\author[orcid=0000-0001-8928-4465]{Andrea~Weibel}
\affiliation{Department of Astronomy, University of Geneva, Chemin Pegasi 51, 1290 Versoix, Switzerland}
\email[]{andrea.weibel@unige.ch}

\author[orcid=0000-0003-3997-5705]{Rohan~P.~Naidu}
\affiliation{MIT Kavli Institute for Astrophysics and Space Research, 70 Vassar Street, Cambridge, MA 02139, USA}
\email[]{rnaidu@mit.edu}

\author[orcid=0000-0002-0463-9528]{Yilun~Ma}
\affiliation{Department of Astrophysical Sciences, Princeton University, Princeton, NJ 08544, USA}
\email[]{yilun@princeton.edu}

\author[0000-0001-9978-2601]{Aidan~P.~Cloonan}\thanks{NSF Graduate Research Fellow}
\affiliation{Department of Astronomy, University of Massachusetts, Amherst, MA 01003, USA}
\email[]{apcloonan@umass.edu}

\author[orcid=0009-0008-9877-5512]{Aayush~Desai}
\affiliation{Institute of Science and Technology Austria (ISTA), Am Campus 1, 3400 Klosterneuburg, Austria}
\email{aayush.desai@ist.ac.at}

\author[orcid=0000-0002-2380-9801]{Anna de Graaff}\thanks{Clay Fellow}
\affiliation{Center for Astrophysics, Harvard \& Smithsonian, 60 Garden St, Cambridge, MA 02138, USA}
\affiliation{Max-Planck-Institut f\"ur Astronomie, K\"onigstuhl 17, D-69117 Heidelberg, Germany}
\email[]{degraaff@mpia.de}

\author[orcid=0000-0002-5612-3427]{Jenny~E.~Greene}
\affiliation{Department of Astrophysical Sciences, Princeton University, Princeton, NJ 08544, USA}
\email[]{jgreene@astro.princeton.edu}

\author[0000-0002-8896-6496]{Christian~Kragh~Jespersen}
\affiliation{Department of Astrophysical Sciences, Princeton University, Princeton, NJ 08544, USA}
\email[]{cj1223@princeton.edu}

\author[orcid=0000-0001-5346-6048]{Ivan~G.~Kramarenko}
\affiliation{Institute of Science and Technology Austria (ISTA), Am Campus 1, 3400 Klosterneuburg, Austria}
\email{ivan.kramarenko@ist.ac.at}

\author[orcid=0000-0002-9572-7813]{Sara~Mascia}
\affiliation{Institute of Science and Technology Austria (ISTA), Am Campus 1, 3400 Klosterneuburg, Austria}
\email{sara.mascia@ist.ac.at}

\author[orcid=0000-0001-5851-6649]{Pascal~A.~Oesch}
\affiliation{Department of Astronomy, University of Geneva, Chemin Pegasi 51, 1290 Versoix, Switzerland}
\affiliation{Cosmic Dawn Center (DAWN), Copenhagen, Denmark}
\affiliation{Niels Bohr Institute, University of Copenhagen, Jagtvej 128, K{\o}benhavn N, DK-2200, Denmark}
\email[]{pascal.oesch@unige.ch}

\author[orcid=0009-0007-3791-7890]{Wendy~Q.~Sun}
\affiliation{MIT Kavli Institute for Astrophysics and Space Research, 70 Vassar Street, Cambridge, MA 02139, USA}
\email{wendysun@mit.edu}

\author[orcid=0000-0003-2919-7495]{Christina~C.~Williams}
\affiliation{NSF–DOE Vera C. Rubin Observatory/NSF NOIRLab, 950 N. Cherry Ave., Tucson, AZ 85719, USA}
\affiliation{Steward Observatory, University of Arizona, 933 N. Cherry Ave., Tucson, AZ 85721, USA}
\email[]{christina.williams@noirlab.edu}

\begin{abstract}

Recent studies at high redshift have revealed an enigmatic class of Little Red Dots (LRDs) with extreme Balmer breaks, stronger than in any stellar atmosphere. However, it is unclear whether such objects exist at lower redshift, especially given the low number of LRDs reported at $z\lesssim 2$.
Here we report the discovery of PAN-BH*-1, an LRD with an extreme Balmer break at $z=1.73$, identified from JWST/NIRCam pure-parallel imaging taken by the PANORAMIC survey, and confirmed by deep VLT/X-Shooter spectroscopy. The rest-optical to near-infrared spectral energy distribution of PAN-BH*-1 is consistent with a photospheric continuum with effective temperature $T_{\rm eff}\approx 4800$~K. The broad H$\alpha$ emission line shows remarkably deep absorption, stronger than previously measured in any LRD. The absorption trough spans from $-520$~km/s to $+267$~km/s with respect to the systemic redshift. The presence of blue- and red-shifted absorption suggests complex dynamics of the obscuring gas along the line of sight. We speculate that the absorption trough can be produced by a thick wind launched from a thick, rotating photospheric disk, the latter being the source of the red optical continuum.
While the source is unresolved in the rest-optical JWST data ($r_{\rm eff}<47$~pc), the rest-NUV HST imaging shows an extended morphology with $r_{\rm eff}=1.0^{+0.5}_{-0.3}$~kpc, that we interpret as a host galaxy with a stellar mass $\sim 10^8$~$M_\odot$, in line with the narrow H$\alpha$ emission. The discovery of this object at cosmic noon highlights the feasibility of systematic searches for extreme LRDs with wide-area facilities such as Euclid and Roman.

\end{abstract}


\section{Introduction}

The unprecedented sensitivity of {\it JWST} has enabled the discovery of a new, abundant population of objects at redshifts $z\approx3-9$ nicknamed the ``Little Red Dots'' (LRDs). These are characterized by their compact rest-frame optical morphology, broad emission lines, and a characteristic rest-UV to optical ``V-shape'' in their spectral energy distributions \citep[SED; e.g.,][]{kocevski20231, matthee2024, kokorev20242,Labbe2025}.

The nature of LRDs is highly debated \citep[see][for a recent overview]{inayoshi20252} as the LRDs show systematic differences with respect to other types of active galactic nuclei (AGN), such as faintness in X-rays \citep[e.g.,][]{ananna2024, yue20241}, mid-to-far infrared dust emission \citep[e.g.,][]{leung2024, williams2024, setton2025, xiao2025, Delvecchio2025}, and radio \citep[e.g.,][cf. \citealt{gloudemans2025}]{latif2025, perger2025, mazzolari2024}. 

A recurring spectral feature of LRDs is the presence of a strong Balmer break \citep[e.g.,][]{wang2024,setton20241, hviding2025,sun2026}, in some cases stronger than any star or stellar population can produce.  The two most prominent examples known to date are \cliff{} at $z\approx 3.5$ \citep{degraaff20255} and \mombhs{} at $z\approx 7.8$ \citep{naidu2025}. The joint appearance of strong Balmer lines as well as strong Balmer breaks has been modeled as being due to absorption by a dense, neutral gas with a high column density in the line of sight to a highly ionizing source \citep{inayoshi20253, ji2025,torralba2025,sneppen2026}. These observations have sparked the development of new theoretical models, ranging from a spherical envelope analogous to stellar atmospheres \citep[e.g.,][]{liu2025, kido2025, begelman2025, nandal2026} or a thick accretion disk \citep[e.g.,][]{liu2025, Liu2026, inayoshi20251, chen2026}.

Besides their spectral features, the evolution of the LRD number densities is also in stark contrast to other types of AGNs \citep[e.g.,][]{Inayoshi25_zevo}. At $4 \lesssim z \lesssim 7$, LRDs represent a few percent of the galaxy population \citep[e.g.,][]{kocevski20231, kocevski2024, maiolino20241, greene2024, matthee2024, kokorev20242,lin2024}, with number densities of $\gtrsim$$10^{-5}$~\unit{Mpc^{-3}}. The number density does not appear to drop quickly beyond $z>5$ \citep[e.g.,][]{zhang20252}, with various LRDs having been confirmed at $z>8$ \citep{kokorev2023,taylor2025,Tripodi25Nat}, well beyond the quasar redshift record \citep[][]{Wang2021_z7q}. Photometric LRD candidates exist beyond $z>10$ \citep{Tanaka25}.
In turn, the number density of LRDs seems to decline steeply at $z<4$ \citep[e.g.,][]{ma20251}, with estimates of a number density of $\sim10^{-6}$ cMpc$^{-3}$ at $z\sim2$ and even $\sim10^{-10}$ cMpc$^{-3}$ at $z\approx0.3$ \citep{lin20252}. While it is challenging to ensure a uniform selection function across such a large redshift baseline and dedicated spectroscopic follow-up of such lower redshift candidates has only just started, it is challenging to attribute five orders of magnitude to such effects.

Motivated by the discovery of rare objects with extreme Balmer breaks at $z>3$ and the very small number of known LRDs at lower redshift, we performed a dedicated search for extreme Balmer break objects using a template-match approach on a large compilation of JWST NIRCam data over $\approx0.3$ deg$^2$ and $z\approx1.5-7.0$. This survey will be presented in~A.~Weibel et al. (in prep.). As part of an ongoing ground-based spectroscopic campaign of LRD candidates at $z\sim2$ \citep{ma20251}, we followed up the most luminous candidate with a photometric redshift of $z\approx2$ with the \mbox{\it X-Shooter} spectrograph on the \textit{Very Large Telescope} (VLT). In this letter, we present the discovery and spectroscopic confirmation of \panbhs{}, a luminous LRD at $z=1.73$ with an extreme Balmer break comparable to the strongest observed in any LRD (and, in general, any astrophysical source). The low redshift of this source enables high-resolution spectroscopy from ground-based observatories that is otherwise impossible to obtain at high redshift.

Throughout this letter we use a $\Lambda$CDM cosmology with $\Omega_m = 0.31$, $\Omega_\Lambda = 0.69$ and $h = 0.677$ as described by~\citet{collaboration2020}. All the magnitudes are given in the AB system \citep{Oke83}.

\section{Observations}\label{sec:obs}

\begin{figure*}
    \centering
    \includegraphics[width=\linewidth]{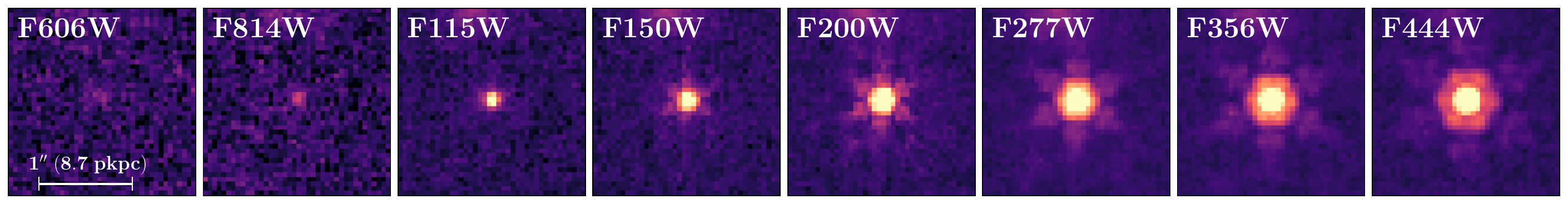}
    \includegraphics[width=0.85\linewidth]{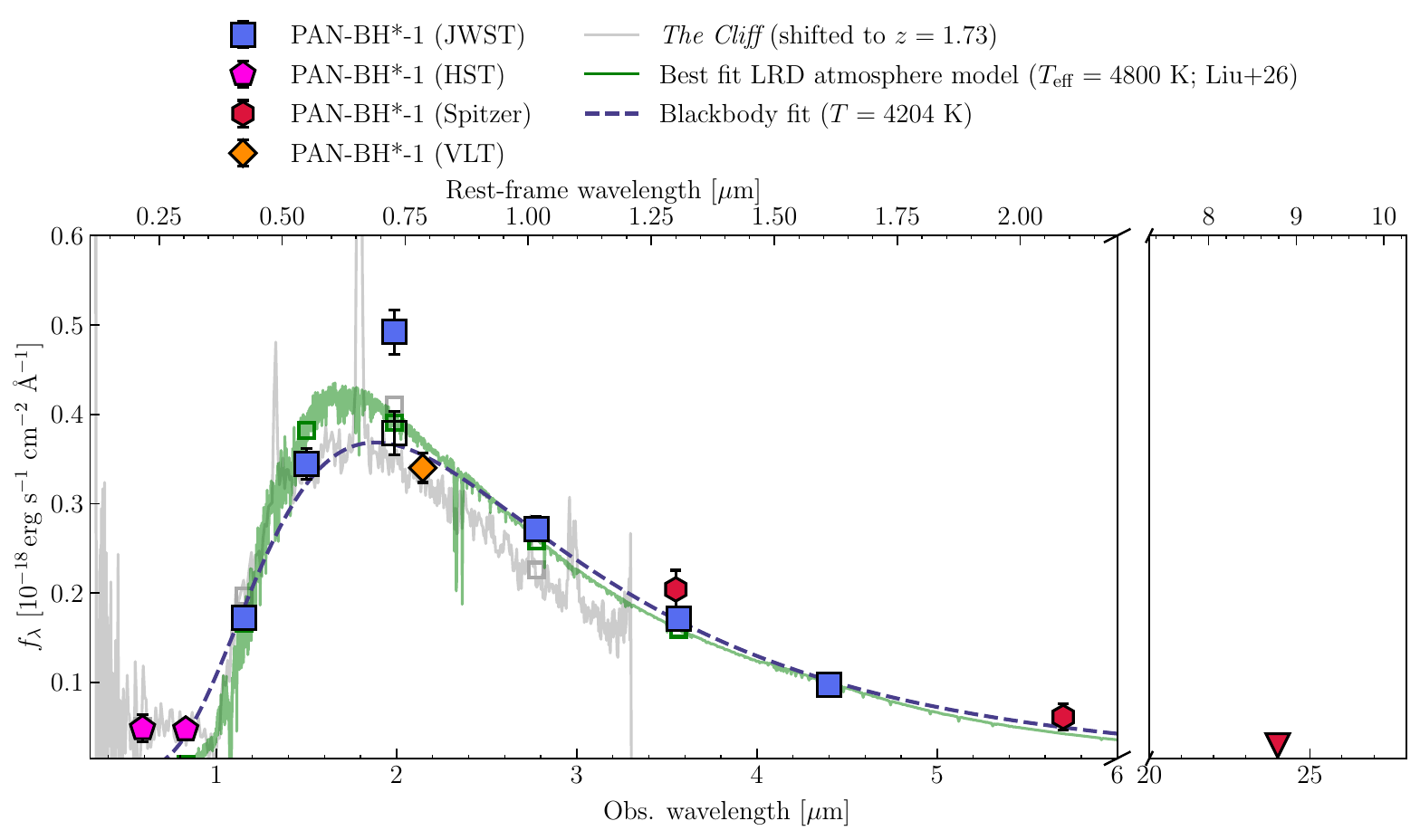}
    \caption{{\bf SED of \panbhs{}} {\bf Top:} Cutouts from all the HST and JWST images in which \panbhs{} is covered. It shows a remarkably compact morphology in all the wavelengths, resolved only in the HST F606W and F814W bands (Sect.~\ref{sec:morph}). {\bf Bottom:} Photometry from JWST/NIRCam (blue squares), HST/ACS (purple pentagons), and Spitzer/IRAC+MIPS (red hexagons, and red triangle for the 5$\sigma$ upper limit). The empty square is the F200W flux after subtracting the \Halpha{} flux measured from X-Shooter spectroscopy. We show the spectrum of \cliff{} for comparison (gray line), shifted to $z=1.73$ and normalized to the F150W flux of \panbhs{}. We also show the best-fitting Blackbody spectrum (blue dashed line) and the best model from the synthetic LRD atmosphere models from~\citet{Liu2026}, shifted to $z=1.73$ (green line), under-sampled a factor 500 for clarity.}
    \label{fig:SED}
\end{figure*}

\begin{figure}
    \centering
    \includegraphics[width=\linewidth]{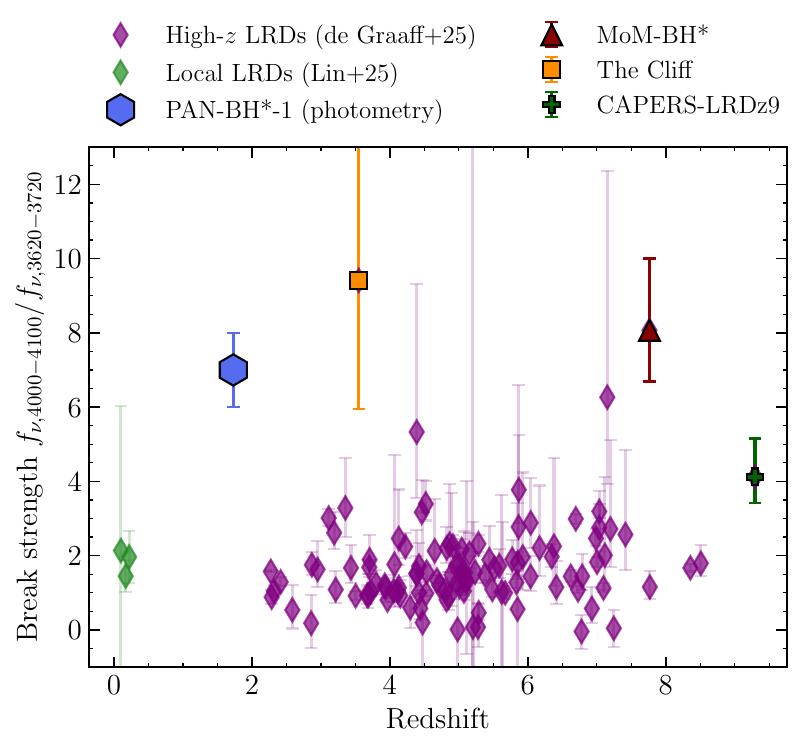}
    \caption{{\bf Spectroscopic sample of LRDs by redshift and Balmer break strength.} We plot the redshift and Balmer break strength of \panbhs{}, and the JWST sample from~\citet{degraaff20251} (purple diamonds), and three local LRDs in~\citet{lin20252}, for comparison. We also highlight three sources with a particularly strong Balmer break: \cliff{} \citep[][]{degraaff20255}, \mombhs{} \citep[][]{naidu2025}, and \caperslrd{} \citep{taylor2025}. The Balmer break strength of the JWST spectroscopic sample is computed as $f_{\nu, 4000-4100}/f_{\nu, 3620-3720}$, whereas the value for \panbhs{} is directly obtained from the F115W/F814W photometry.}
    \label{fig:bbreak_z}
\end{figure}

\begin{figure}
    \centering
    \includegraphics[width=\linewidth]{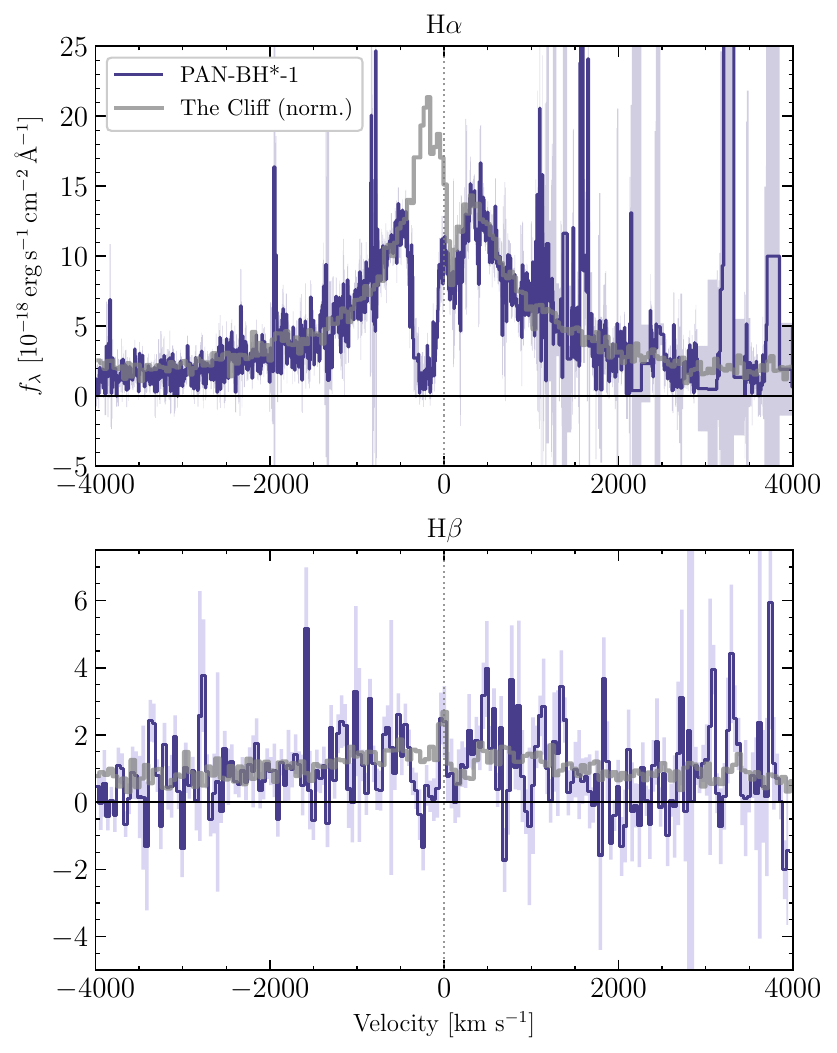}
    \caption{{\bf X-Shooter spectrum of \Halpha{} and \Hbeta{} of \panbhs{}} (blue). We compare to the spectrum of \cliff{} (gray; data from JWST DDT \#9433), normalized in each panel to the flux of \panbhs{} in the range $v\in (-3000, -2000)~{\rm km\,s^{-1}}$. Due to the low S/N, the \Hbeta{} spectrum of \panbhs{} is re-binned to a coarser grid by a factor 5, after masking the most relevant sky lines.}
    \label{fig:Halpha_vs_cliff}
\end{figure}

\subsection{Photometry and source selection}

We identified \panbhs{} (ID: PAN-1115, RA, DEC: $40.015835$, $-1.659363$ J2000) as part of a systematic search across $\approx0.3$ deg$^{2}$ of JWST NIRCam legacy imaging comprising at least six filters of coverage \citep[][]{Weibel25UVLF}. Notably, this dataset includes the Cycle~1 pure parallel survey PANORAMIC \citep[PID: 2514, PIs: Williams \& Oesch;][]{williams2025} that contributes 28 of the 35 independent lines of sight, thereby enabling the discovery of rare objects such as \panbhs{} across diverse large scale structure environments. Specifically, this source was identified in the footprint \texttt{j024000m0142} of the PANORAMIC DR1\footnote{\url{https://panoramic-jwst.github.io/panoramic_dr1_summary_v0.html}}, which is adjacent to the Abell 370 field \citep[][]{abell1989}, where archival images by the \textit{Hubble Space Telescope} (HST) are available from the BUFFALO survey \citep[][]{steinhardt2020}. The HST/ACS images were processed with \texttt{grizli} and also released as part of the PANORAMIC dataset. 

\panbhs{} is in the outskirts of the Abell 370 lensing cluster, but the magnification is only $\mu \approx 1.05$ according to the models from~\citet{Niemiec2023}. Throughout the rest of the paper we report the uncorrected flux measurements, since the effect of magnification ($\sim 5\%$) is negligible given the uncertainties in the observations and the lensing model.

The search strategy and full photometric selection are described in a companion paper (A.~Weibel et al., in prep.). Briefly, that work presents a new selection of LRDs as a combination of a ``black hole star'' template \citep[BH*;][]{naidu2025} embedded in a host galaxy, instead of typically used ``V-shaped" selections \citep[e.g.,][]{kokorev20242, kocevski2024}. The host galaxies are modeled using \texttt{eazy}'s \texttt{blue\_sfhz} templates. The BH*s are modeled using a novel template set comprising empirical luminosity-based stacks constructed in \citet{sun2026}, the \texttt{cloudy} template from \citet{naidu2025}, and by using spectra of prominent LRDs spanning the observed effective temperature range \citep[][]{labbe2024, degraaff20255, Wang26}. 

\panbhs{} stood out as one of the few sources where the BH* template effectively dominated all the light over the full wavelength range covered by NIRCam (hence the name).
The redshift of \panbhs{} was estimated to be $z_{\rm phot}=1.85$. Follow-up VLT/X-Shooter spectroscopy confirmed the redshift as $z_{\rm spec}=1.731$ (see Sect.~\ref{sec:ha_fit}).

\panbhs{} is also covered by archival data from the VLT with the HAWK-I camera in the $K_s$ band \citep{Brammer16} and in data from the {\it Spitzer Space Telescope} in IRAC bands 1 and 3 (3.6~\unit{\mu m} and 5.7~\unit{\mu m}), and MIPS 24~$\micron$ \citep{Capak2019}.  \panbhs{} is detected in the $K_s$ band and in the two IRAC filters. Performing {Spitzer} photometry of this source is challenging due to the large PSF and a neighbouring source, especially in the MIPS band. However, the NIRCam photometry of the neighbouring source suggests it has a limited contribution to the IRAC fluxes. The details of the photometry extraction are described in Appendix~\ref{sec:spitzer_phot}, and the measured magnitudes in Table~\ref{tab:pan1115_photometry}.

\subsection{VLT/X-Shooter spectroscopy}

\panbhs{} was observed for 5.8 ks with the X-Shooter spectrograph \citep{vernet11} on the VLT as a bright backup target for program 116.294D (PI: Matthee) in visitor mode on 17 December 2025. The main aim of this program was to confirm candidate LRDs at cosmic noon \citep{ma20251}. These observations confirmed the redshift through detection of H$\alpha$ at $z=1.731$. A DDT program (ID 116.2AQ0; PI: Matthee) obtained additional follow-up data of \panbhs{} in service mode for 26.2~ks during 10--26 January 2026, yielding a total exposure time of 8.9 hours. \mbox{X-Shooter} observes with three arms simultaneously, UVB, VIS and NIR, covering rest-frame wavelengths $\approx0.14$--$0.9$~$\mu$m, albeit hampered by skyline emission and telluric absorption, primarily in the rest-frame optical. 

The observing conditions were clear, with a seeing ranging from 0.5--0.7\arcsec{} (median 0.6\arcsec{}). The service mode observations were primarily conducted during dark nights, with some grey ($\rm FLI = 0.03$--0.6, median 0.1), and a typical airmass of 1.35. We used UVB, VIS, and NIR slits with widths 1.0\arcsec{}, 0.9\arcsec{}, and 0.9\arcsec{}, yielding a nominal resolution of $R=5400$, 8900, and 5600, respectively (FWHM $\sim$53~\unit{km.s^{-1}} for NIR). The target acquisition was done using blind offsets from a reference star, due to the target being too faint for direct acquisition. We used a standard nodding on the slit pattern, with 4\arcsec{} nod throws in an $ABBA$ pattern, and 1\arcsec{} jitters in the NIR arm to improve the sky subtraction. In each observing block of $\approx1$ hour, the exposure times were 700, 655 and ($2\times$)365~s for the three arms at each nod position.

The reduction of the X-Shooter data uses a combination of \texttt{esorex} libraries\footnote{\url{https://www.eso.org/sci/software/cpl/esorex.html}} and python code based on the reduction pipeline employed in \cite{matthee21}. Each observing block was reduced separately. We used standard stars taken during the observing night for a first-pass flux calibration. Telluric corrections were applied using the molecfit tool \citep[][]{molecfit} implemented in the X-Shooter \texttt{esorex} pipeline. Telluric stars were observed during the visitor nights, but they were not always observed during the service mode observations in January. For those observations, we took the telluric star that was observed at the closest observing date. Based on the variation in telluric absorption among the reference stars taken during this period, we estimate the variation in the transmission and propagate the uncertainty in the telluric correction. For each observing block, we then extracted an optimally-extracted 1D spectrum using the spatial profile of the H$\alpha$ line, thus accounting for seeing variations and (more importantly) minor errors of the accuracy of the slit pointing. Before median-combining these spectra, we normalize them by the median H$\alpha$ flux of all observations to account for variations in slit losses and flux calibrations.

Besides H$\alpha$ (integrated $\rm S/N=75$) and H$\beta$ (integrated $\rm S/N = 6$; Sect.~\ref{sec:ha_fit}), we also detect continuum emission in the best regions in the $H$ and $K$ bands at $1.6\ \mu$m and $2.1\ \mu$m, respectively, with a low S/N $\sim1$ per resolution element. Unfortunately, the [O\,{\sc iii}] $\lambda\lambda{4960,5008}$ doublet is undetectable because the observed wavelengths are impacted by very strong telluric absorption. No other lines or continuum are detected in the X-Shooter spectrum.

\section{Properties of \panbhs{}}\label{sec:properties}

\subsection{Spectral shape: a photospheric continuum with strong \Halpha{} emission}\label{sec:sed}

The photometric SED of \panbhs{} has remarkable similarities with \cliff{} (Fig.~\ref{fig:SED}): luminous in the rest-optical, with a sudden drop towards the rest-UV around the Balmer limit, and very weak near-to-mid infrared continuum emission. 
With a rough extrapolation of the two HST photometric points using a powerlaw fit ($f_\lambda \propto \lambda^{\beta}$), we obtain a UV slope of $\beta = -0.1 \pm 1.2$, and $M_{\rm UV} = -16.7 \pm 0.7$.
For the rest-frame optical to NIR data, we fit a Planck blackbody law to the JWST data points, after subtracting the measured \Halpha{} flux (see Sect.~\ref{sec:ha_fit}) from the F200W photometry. The rest-optical and NIR photometry of \panbhs{} is remarkably well described by a single temperature blackbody with $T=4204$~K (with a best-fit $\chi_\nu^2 = 1.1$).  We measure the strength of the Balmer break from the $f_\nu$ ratio $\rm F115W/F814W = 7 \pm 1$, in line with the Balmer break strengths of \cliff{}\  \citep[$6.9^{+2.8}_{-1.5}$;][]{degraaff20255}\footnote{This value is slightly different to the one reported in~\citet{degraaff20251}, due both works using different data reduction versions.} and \mombhs{} \citep[$7.8\pm1.8$;][]{naidu2025}, measured from JWST/NIRSpec PRISM spectra as $f_{\nu, 4000-4100}/f_{\nu, 3620-3720}$. In Fig.~\ref{fig:bbreak_z} we compare the Balmer break strength with the spectroscopic sample of~\citet{degraaff20251}, showing that out of 134 sources, only two have breaks significantly above 5. This suggests that \panbhs{} has among the most extreme Balmer breaks known, although we caution that our value is derived from wide-band photometry with pivot wavelengths corresponding to 4212~\AA{} and 3042~\AA{}, respectively, rather than from spectroscopy.

\subsection{\Halpha{} and \Hbeta{} emission lines}\label{sec:ha_fit}

\begin{figure*}
    \centering
    \includegraphics[width=0.7\linewidth]{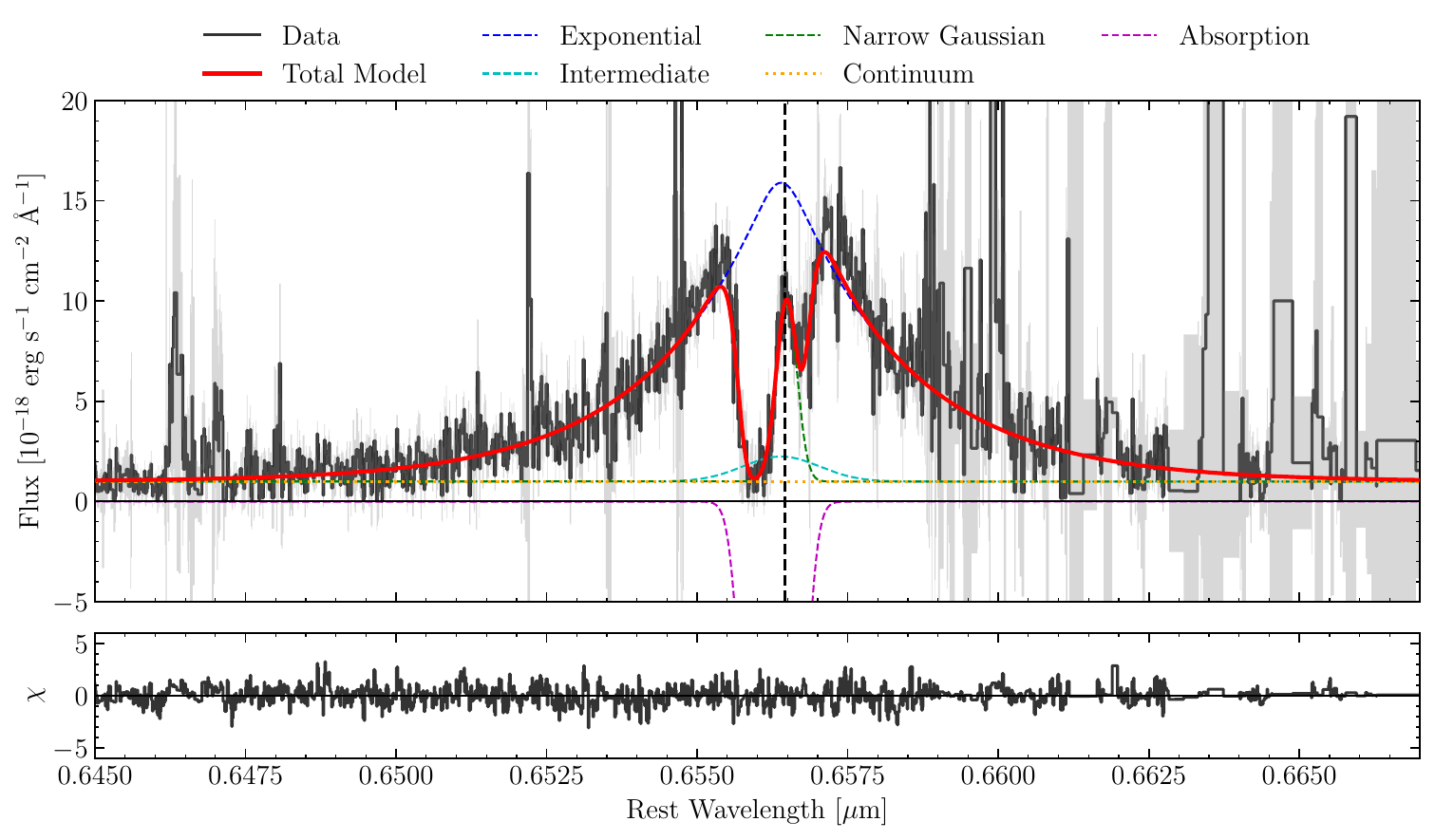}
    \caption{{\bf \Halpha{} spectrum, and best fit to our fiducial model.} We show the X-Shooter $R\sim 5600$ spectrum of the \Halpha{} line of \panbhs{}, along with the best-fit to the model described in Sect.~\ref{sec:ha_fit}; total model (red solid line) and individual components (discontinuous colour lines). The red wing of the line is severely affected by telluric absorption, thus the large uncertainties.}
    \label{fig:halpha_fit_full}
\end{figure*}

\begin{figure}
    \centering
    \includegraphics[width=\linewidth]{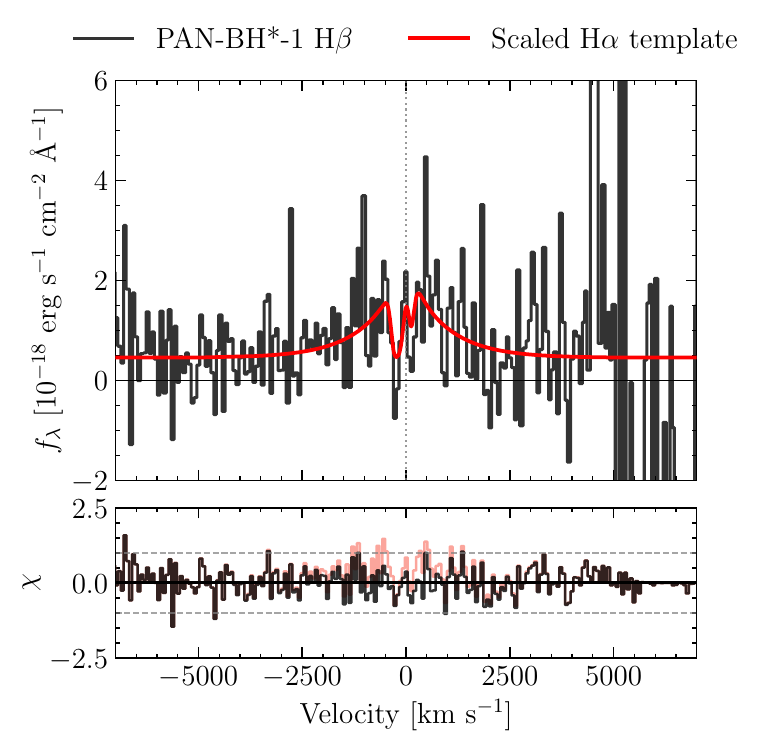}
    \caption{{\bf \Hbeta{} spectrum.} The spectrum is rebinned a factor 10 with inverse variance flux weighting for visual clarity, due to the low S/N. We compare to the best-fit \Halpha{} model, scaled by a factor 0.112. In the bottom panel we show the $\chi$ residuals between the spectrum and the re-scaled \Halpha{} model in black, and for only continuum in pink ($\rm\Delta BIC = 121$ strongly favouring the presence of a broad \Hbeta{} line).}
    \label{fig:hb_spectrum}
\end{figure}

\begin{table}
\centering
\caption{Properties of \panbhs{}.}
\label{tab:properties}
\begin{tabular}{lcc}
\hline
Parameter & Value & Unit \\
\hline
\multicolumn{3}{l}{{\bf Width (FWHM; \Halpha)}} \\
\hline
Exponential    & $1257 \pm 27$ & km s$^{-1}$ \\
Intermediate & $687 \pm 43$ & km s$^{-1}$ \\
Narrow       & $184 \pm 12$ & km s$^{-1}$ \\
Absorption   & $283 \pm 8$ & km s$^{-1}$ \\
\hline
\multicolumn{3}{l}{{\bf Flux (\Halpha{})}} \\
\hline
Exponential & $643 \pm 7$ & $10^{-18}$ erg s$^{-1}$ cm$^{-2}$ \\
Intermediate & $19 \pm 5$ & $10^{-18}$ erg s$^{-1}$ cm$^{-2}$ \\
Narrow       & $38 \pm 3$ & $10^{-18}$ erg s$^{-1}$ cm$^{-2}$ \\
Total & $522 \pm 7$ & $10^{-18}$ erg s$^{-1}$ cm$^{-2}$ \\
\hline
\multicolumn{3}{l}{{\bf General properties}} \\
\hline
$\log_{10}$($L_{\mathrm{H}\alpha}\rm /erg s^{-1}$) & $43.046 \pm 0.006$ & --\\
$\rm EW_0(\Halpha)$ & $520\pm20$ & \AA{} \\
SFR(\Halpha{}, narrow)$^\dagger$ & $2.1 \pm 0.2$ & $M_\odot\,\rm yr^{-1}$ \\
SFR(\Halpha{}, narrow)$^\star$ & $3.3 \pm 0.3$ & $M_\odot\,\rm yr^{-1}$ \\
$r_{\rm eff, UV}$ (F606W+F814W) & $1.0^{+0.5}_{-0.3}$ & kpc \\
$r_{\rm eff, opt}$ (F200W) & $<0.047$ & kpc \\
$\Halpha{}/\Hbeta{}$ (Total) & $> 9.4$ & -- \\
$\Halpha{}/\Hbeta{}$ (Narrow) & $5 \pm 1$ & -- \\
\hline
\end{tabular}
\tablecomments{
$^\dagger$Calibration from \citet{kramarenko2025}.
$^\star$Calibration from \citet{kennicutt20121}. SFR values calculated assuming no dust attenuation.
}
\end{table}

The \Halpha{} profile appears as a complex combination of a broad line with strong absorption close to the systemic redshift. We fit the \Halpha{} emission line with a similar model as the one used in~\citet{torralba2025} and \citet{Matthee2026}. The \Halpha{} model consists of two Gaussian emission components (with narrow and intermediate line width), and a broad symmetric exponential convolved with the intermediate profile and parametrized as in~\citet{deugenio20252}. The absorption is implemented as an opacity law defined as $e^{-\tau(\lambda)}$, where $\tau(\lambda)$ also follows a single Gaussian velocity distribution \citep{deugenio20251, deugenio20252, torralba2025}. For simplicity, we assume a covering factor of $C_{\rm f}=1$ for the absorbing gas. In previous works, the width of the narrow component is tied to that of \oiii{}, assuming both components come from the same region, often interpreted as the interstellar medium (ISM) of the host galaxy. In this case, we have no information about \oiii{} due to this doublet falling in a wavelength range heavily affected by strong telluric absorption. We fit the \Halpha{} line after masking relevant sky lines lines and strong telluric absorption bands. The fitted \Halpha{} parameters are listed in Table~\ref{tab:properties} and the best-fit model is shown in Fig.~\ref{fig:halpha_fit_full}. The absorption feature is notably strong, with an equivalent width of $\rm EW_{\rm abs} = -148 \pm 12$~\AA{} with respect to the fitted continuum and $12.2\pm 0.2$~\AA{} if including the broad emission component. The absorption corresponds to a Balmer optical depth at the line center of $\tau_0 = 9.6^{+1.6}_{-1.1}$, reaching roughly the continuum level. The FWHM of the single Gaussian fitted to the absorber is $283\pm 8$~\unit{km.s^{-1}}, and is offset from the systemic redshift by $-94\pm 4$~\unit{km.s^{-1}}. We note that this parametrization is somewhat arbitrary, and we discuss in detail the absorber properties in Sect.~\ref{sec:abs}.

The \Hbeta{} line is marginally detected. After undersampling the spectrum by a factor 5, a hint of a weak narrow component can be identified (Fig.~\ref{fig:Halpha_vs_cliff}), along with a tentative absorption at the same mean velocity as in \Halpha{}. We fit the best \Halpha{} model to the \Hbeta{} spectrum, only re-scaling it by a multiplicative factor, and adding a flat continuum component. By doing this, we find an \Hbeta{} flux of $(47\pm 8)\times$~\unit{10^{-18}.erg.s^{-1}.cm^{-2}} ($\rm S/N \approx 6$). Conservatively, we obtain a Balmer decrement of $\Halpha/\Hbeta > 9.4$ (at a 3$\sigma$ confidence level), in line with the high decrements found for the LRD population \citep[e.g.,][]{Nikopoulos2025, degraaff20251,Matthee2026}. In Fig.~\ref{fig:hb_spectrum} we show the \Hbeta{} spectrum compared to the re-scaled \Halpha{} model. By matching the best-fit \Halpha{} profile with the data at the expected observed wavelength for \Hbeta{} ($\pm 5000$~\unit{km.s^{-1}}), we obtain a better agreement ($\chi^2_\nu = 1.03$, $\rm BIC=1537$) than fitting a flat continuum only ($\chi^2_\nu=1.12$, $\rm BIC=1658$) with $\rm\Delta BIC=121\gg10$, strongly favouring a detection of a broad \Hbeta{} emission line, and securing the spectroscopic redshift. Similarly, we fit a narrow Gaussian to \Hbeta{} with the same width and velocity as the \Halpha{} best-fit model, assuming a completely saturated absorption. We obtain a Balmer decrement for the narrow component of $\Halpha/\Hbeta = 5 \pm 1$, which would imply a dust extinction of $A_v = 1.9^{+2.4}_{-1.4}$ using a~\citet{cardelli1989} attenuation law, under the assumption of case B recombination. However, due to the low S/N of \Hbeta{} this result is only tentative, and compatible with a standard Case B value within $\sim2\sigma$.

\subsection{Spatial morphology}\label{sec:morph}

In order to assess whether \panbhs{} is spatially resolved, we use the Bayesian profile fitting software \texttt{pysersic} \citep{pysersic}\footnote{\url{https://github.com/pysersic/pysersic}} to fit a single S\'{e}rsic profile to the JWST and HST imaging data of \panbhs{}. For JWST/NIRCam, we choose F200W as the filter with the highest signal-to-noise ratio in the short wavelength channel, benefitting from a high spatial resolution and probing rest-frame optical wavelengths. To model its PSF, we use version 2.2.0 of the \texttt{stpsf} software \citep[formerly \texttt{webbpsf},][]{Perrin2014}. For the two HST bands F606W and F814W, we instead construct empirical PSFs from public imaging data in the GOODS-S field following \citet{weibel2024}. In all three bands, we sample the posterior with the \textit{No U-turn} sampler in two chains with 1000 warm-up and 2000 sampling steps each. 
We find that \panbhs{} is unresolved with NIRCam in F200W where the effective radius converges towards the edge of the prior at 0.25 pixels. Using the 95th percentile of the posterior chains as an upper limit on the effective radius, we find a rest-optical size of $r_{\rm eff}<47$~pc.

\panbhs{} appears to be resolved in the HST images corresponding to rest-frame pivot 0.2~\unit{\mu m} and 0.3~\unit{\mu m}, respectively. Due to the low signal-to-noise of the F606W and F814W photometry, we fit both bands simultaneously fixing all the morphological parameters in both images. We measure physical effective radii of $r_{\rm eff}=1.0_{.-0.3}^{+0.5}$~kpc (see Appendix~\ref{sec:morph_app}). The modest stretching by the foreground Abell 370 lensing cluster could imply a correction of $\sim 10\%$ to the measured radius \citep[][]{Niemiec2023}, which we disregard given the uncertainties. These measured sizes are consistent with the typical sizes for galaxies with a stellar mass $\lesssim\,10^9$~$M_\odot$ at $z=1.75$ \citep{vanderwel2014}. These findings are consistent with the scenario of a compact LRD ``engine'' dominating the rest-optical light embedded in a host galaxy, whose contribution becomes significant blueward of the Balmer break \citep[see][for a relevant discussion]{Cloonan2026}.

\section{Absorber kinematics}\label{sec:abs}

As described in Sect.~\ref{sec:ha_fit}, the velocity distribution of the absorber is empirically modeled with a Gaussian, which we find has a central velocity of $-94 \pm 4$~\unit{km.s^{-1}} relative to redshift of the narrow emission component (adopted as systemic). The absorption trough extends from negative to positive velocities with respect to the redshift of the narrow component, but also with respect to the center of the symmetric exponential wings. However, there are several degeneracies between the shape of the absorber and other components of the emission line, such as the narrow central emission (see Sect.~\ref{sec:ha_fit}). Furthermore, direct interpretation of the absorber center velocity shift is challenging in an optically thick gas with presumably complex dynamics, and it does not necessarily trace bulk motion. A more robust, physically motivated pair of quantities, are the minimum and maximum absorber velocities. We define them as the values where the transmission of the Balmer absorber increases to 99\%, $v_{99}^{\rm blue} = -520 \pm 13$~\unit{km.s^{-1}} and $v_{99}^{\rm red} = 267 \pm 11$~\unit{km.s^{-1}}. These values trace the largest velocities in the line of sight of gas with significant Balmer absorption. The absorbing trough extends over $787 \pm 17$~\unit{km.s^{-1}} under this definition. The values of $v_{99}^{\rm blue}$ and $v_{99}^{\rm red}$ are relatively agnostic to the choice of the shape of the absorber, since they are determined by the wavelength where the line profile deviates from a broad, symmetric exponential profile.
In Fig.~\ref{fig:wind_cartoon} we illustrate three proposed configurations of the velocity distribution of the absorbing gas that could explain the shape of the observed Balmer absorption, and we discuss these scenarios below.

\subsection{Unstable gas flows?}

The fact that there is significant absorption at both negative and positive velocities with respect to the systemic redshift cannot be simply explained by an axisymmetric outflowing or inflowing wind. In the case of observing a compact object through a spherically symmetric, non-turbulent bulk flow, a classical P Cygni profile is expected, with a purely blueshifted absorption (or redshifted if the wind is infalling). The fact that we also see redshifted absorption rules out this simple scenario. In principle, turbulent motions could also produce broadening of the absorbing medium (scenario \textit{a} in Fig.~\ref{fig:wind_cartoon}). However, the required turbulent velocity dispersion $\sigma_{\rm turb} \approx 120$~\unit{km.s^{-1}} (from the Gaussian fit in Sect.~\ref{sec:ha_fit}) is comparable to the mean velocity of the absorption trough, meaning that turbulence dominates the gas flow. In such a case, strong variability of the absorption profile would be expected, given the typical dynamical crossing times \citep[see Sect. 4.1 in][]{deugenio20251}. For example, for a radius of $10^{16}$~cm \citep[e.g.,][]{torralba2025} and a mass of $10^6$~$M_\odot$, the dynamical free fall time is $t_{\rm ff} = \sqrt{\frac{R^3}{GM}}\approx 2.75$~yr. 
Moreover, the turbulent velocity would be highly supersonic, and the dissipation timescale would be comparable to the dynamical time \citep[e.g.,][]{MacLow1999}.
Alternatively, in the context of a strong Balmer absorber at $z\sim7$, \citet{deugenio2026_jadesbt} recently discussed a ``breathing mode'' scenario with cyclic inflows and outflows along the same line of sight, with the gas being in different phases at different depths \citep[scenario \textit{b} in Fig.~\ref{fig:wind_cartoon}; see also][]{Park2017}. In this case, the same arguments regarding the stability of the absorber would apply, and absorber variability is expected on observed timescales of $\sim5$ yr (for a source at $z=1.7$), which is testable with future observations.

\subsection{The case for the disk wind hypothesis}\label{sec:diskwind}

\begin{figure*}
    \centering
    \includegraphics[width=0.8\linewidth]{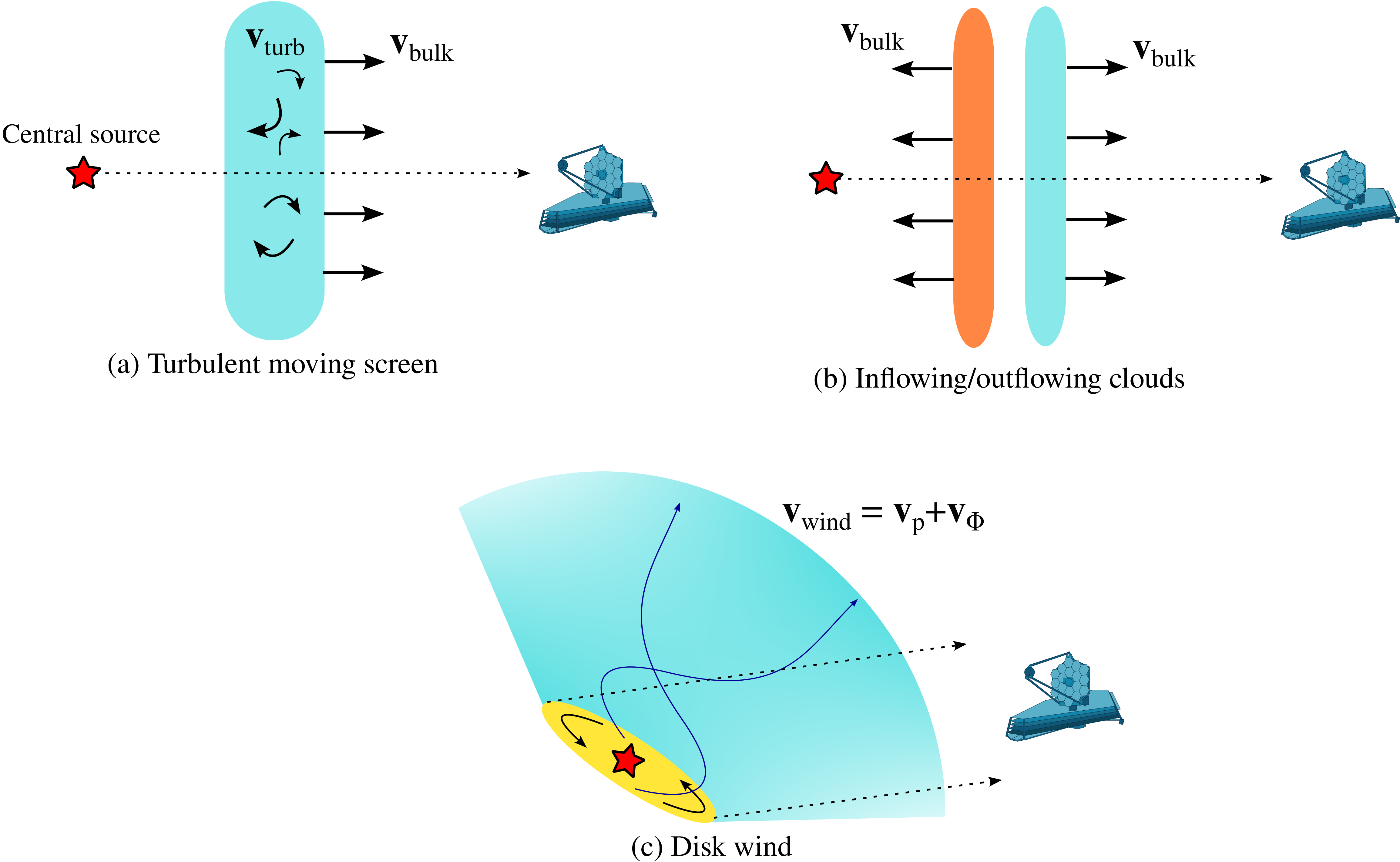}
    \caption{{\rm Geometric configurations for the absorber.} We illustrate three scenarios that could give rise to the observed Balmer absorption in \panbhs{}. In scenario (a), the obscuring agent is a thick screen of gas with a certain bulk velocity, and turbulent motions produce the broadening of the absorption trough. In (b), there are two (or more) absorbers with opposite velocities in the line-of-sight. These first two scenarios are dynamically unstable, therefore variability is expected in the absorption. Lastly, in (c) we observed an extended source through a disk wind with a rotational component ($\mathbf{v}_\phi$) in addition to the poloidal (non-azimuthal) velocity ($\mathbf{v}_p$). In the last scenario, the redshifted absorption is produced by stream lines that oppose the observer when projected along the line-of-sight, despite the fact that the gas is outflowing from the central source.}
    \label{fig:wind_cartoon}
\end{figure*}

An alternative, dynamically stable scenario is a disk wind configuration (scenario \textit{c} in Fig.~\ref{fig:wind_cartoon}).
Here, the wind would be launched from a thick disk near the central engine which we speculate could be the source of the optical continuum emission \citep[e.g.,][]{liu2025, Liu2026, chen2026, zwick2025}. A rotating disk would imprint to the wind an azimuthal velocity component ($\mathbf{v}_\phi$).
Observations at specific lines of sight, particularly for high inclination angles (close to edge-on) where the rotational component dominates the poloidal velocity, can give rise to both blueshifted and redshifted absorption features \citep{Proga2000, Hall2002, hall2013, Proga2004, Giustini2012}.
Most observed LRDs have blueshifted P Cygni-like absorbers \citep[][]{Matthee2026}, that can be naively interpreted as a uniformly expanding shell. The low incidence of redshifted Balmer absorbers in LRD spectra \citep[e.g.,][]{degraaff20255, labbe2024, deugenio20251, deugenio2026_jadesbt, ma20251} can therefore be explained by the requirement of high inclination angles to observe such features \citep[see also][]{sneppen2026}. Such a picture is broadly in line with disk wind models for AGN with broad absorption lines \citep[e.g.,][]{Hall2002, Zhou2019} and around stars with circumstellar disks \citep[e.g.,][]{Erkal2022}, such as accreting T Tauri stars \citep[][]{Edwards2006} or cataclysmic variables \citep[][]{2003ApJ...592L...9P}.

\subsection{Implications of rotating winds for the emission lines of LRDs}

The disk wind hypothesis would imply that a photosphere in the shape of a rotating disk is the source of the optical continuum emission, and drives winds that can explain the observed absorption trough. Emission lines originating in a thin rotating disk would have a double-peaked profile in the idealized case (for most inclination angles), but this is not necessarily true if the disk is not sufficiently thin \citep[e.g.,][]{Murray1995}, for instance, in the case of a puffed-up disk associated with super-Eddington accretion \citep[e.g.,][]{Liu2026}. In addition, most line emission would not be produced directly at the base of the disk, but slightly outside \citep[e.g., via collisional cooling or residual recombination;][]{torralba2025}, where the rotational velocity is lower, and the dynamics are complex \citep[e.g.,][]{Shields1977}.

The Balmer lines of most LRDs are dominated by broad, symmetric exponential components that are associated with broadening by electron scattering \citep[e.g.,][]{rusakov2025, Matthee2026}. For \panbhs{}, the \Halpha{} line profile of \panbhs{} is compatible with a broad exponential profile emerging through a dense wind where the absorption trough is produced.
In dense gas with a large column density of neutral hydrogen, and optically thick to Balmer transitions ($N_{{\rm H\,\text{\sc i}, }2s}\gtrsim 10^{14}$~\unit{cm^{-2}}), resonant scattering effects become important. Crucially, resonant scattering impacts \Halpha{} and \Hbeta{} differently \citep[e.g.,][]{chang2026}, hence the 3D radiative transfer and photon redistribution of both lines may produce different profiles (see e.g., Fig.~2 in~\citealt{2003ApJ...592L...9P}). Therefore, the empirical fitting and interpretation of the absorption profiles becomes non-trivial.
Dedicated radiative-transfer modeling is necessary to study such effects, and they can be tested in other emission lines with high optical depth such as He\,{\sc i} $\lambda 10830$~\AA{}, or resonant lines like C\,{\sc iv} $\lambda 1550$.

\section{Implications for the galaxy and black hole masses}\label{sec:host}

\subsection{Properties of the host galaxy}

Assuming that the narrow component of \Halpha{} corresponds to ISM emission in the host galaxy, we compute the associated star formation rate using the local calibration from~\citet{kennicutt20121} and assuming no dust attenuation. We obtain $\rm SFR(\Halpha{}) = 3.3\pm0.3$~\unit{M_\odot.yr^{-1}}. A somewhat lower value of $\rm SFR(\Halpha{}) = 2.1\pm0.2$~\unit{M_\odot.yr^{-1}} is obtained using the high-redshift ($z\gtrsim4$) calibrations in~\citet{kramarenko2025}, which might be more appropriate for a young dwarf galaxy with a bursty star formation history. The star-formation rates are low, but in line with a main-sequence galaxy with $\log_{10}(M_*/M_\odot)\approx 8$ \citep[extrapolating the relation from][]{Speagle2014S}. Assuming zero dust attenuation, the UV absolute magnitude ($M_{\rm UV}=-16.7\pm 0.7$; Sect.~\ref{sec:sed}) would imply $\rm SFR (UV) = 0.18\pm 0.12$~$M_\odot\rm\,yr^{-1}$ \citep[][]{kennicutt20121}. The discrepancy between the UV and \Halpha{} inferred SFR suggests there is some amount of dust attenuation in the host galaxy.

We derive a dynamical mass from the width of the narrow component \Halpha{} line and the estimated UV size as $\log_{10}(M_{\rm dyn} / M_\odot) = 9.9 \pm 0.2$, adopting the empirical virial correction $K(n)K(q)$ from \citet{vanderwel2022}, where $K(n)$ and $K(q)$ are functions of the best-fit ellipticity and Sérsic index (see Appendix~\ref{sec:morph_app}). Adopting a $M_{\rm dyn} / M_*$ factor of 40 as found by~\citet{degraaff2024_dyn} for dwarf galaxies at high redshift, we infer a stellar mass of $\log_{10}(M_* / M_\odot) = 8.3 \pm 0.2$. However, the $M_{\rm dyn} / M_*$ is very uncertain in this regime, and the uncertainty can span over 1~dex \citep{saldanalopez2025}. We advise caution interpreting this result, as there are large uncertainties in the measurements of the narrow \Halpha{} component, the HST morphology and the empirical relations used.

As discussed in Sect.~\ref{sec:abs}, the absorption profile is compatible with broadening by a rotating disk wind, and numerical modeling of such configurations often predicts a narrow component arising from increased transmission due to purely kinematic effects in the wind geometry \citep{Proga2000, Proga2004, 2003ApJ...592L...9P}. This would be an alternative explanation for at least part of the narrow component flux. On the other hand, most LRDs present narrow \oiii{} emission that is often associated with the host galaxy. Indeed, the ionized gas producing \oiii{} emission should have associated emission in the \Halpha{} and higher order Balmer lines. However, constraining this component largely depends on the assumptions on dust attenuation or ISM conditions, and requires very high S/N and resolution data. Deep, space-based follow-up observations of \panbhs{} would be very constraining for the wind kinematics (e.g., by the joint analysis of \Hbeta{}) and to assess whether a narrow component comes from a host galaxy (e.g., by comparing to a narrow \Hbeta{} component or \oiii{} $\lambda\lambda 4960,5008$).

\subsection{Black hole mass from photosphere models}\label{sec:bh_mass}

The general physical setup of LRDs is an open debate, and their masses a major unknown. Due to the multiple differences with respect to the classical AGN population, the validity of standard virial calibrations has been questioned (e.g., \citealt{rusakov2025, greene2026, torralba2025, sneppen2026}, although see e.g., \citealt{brazzini2025, brazzini2026, Scholtz2026} for an alternative interpretation).

One can obtain a mass estimate assuming a system in radiative equilibrium with $L_{\rm bol}/L_{\rm Edd}=1$ \citep[e.g.,][]{Umeda2026}; this yields a total mass of $\approx 10^6$~$M_\odot$, using the bolometric luminosity from integrating the best-fit blackbody in Sect.~\ref{sec:sed}. 
Recently, \citet{Liu2026} developed a synthetic spectral library of LRD atmosphere models. In these models, the density of the photosphere is regulated by the net surface gravity of an optically thick atmosphere, enabling constraints on the mass of the system. We fit the JWST photometry of \panbhs{} using the models from~\citet{Liu2026}, assuming a negligible contribution from a host galaxy to the optical continuum. The best-fit model has effective temperature $T_{\rm eff} = 4800$~K, surface gravity $\log_{10}(g / \rm cm\,s^{-2}) = -1.5$, and metallicity $\log_{10}(Z / Z_\odot) = -1$ ($\chi^{2}_\nu = 1.7$; see Fig.~\ref{fig:SED}).
The best-fit $\log_{10}g$ implies a total mass of the system (BH plus gas) of $\log_{10}(M_{\rm tot} / M_\odot) \approx 5$ \citep[Eq.~6 in][assuming hydrostatic equilibrium]{Liu2026}. For the second and third best fits we obtain $\log_{10}(g / \rm cm\,s^{-2}) = -1$ and $-2$, respectively ($\chi^2_\nu = 2, 3$, respectively; with the same metallicity and effective temperature), which would imply lower limits to the system mass between $\log_{10}(M_{\rm tot} / M_\odot) \gtrsim 6$ and 4.
The bolometric luminosity of \panbhs{} (from the integral of the best-fit green curve in Fig.~\ref{fig:SED}) implies an Eddington luminosity ratio of $L/L_{\rm Edd}\lesssim 13$, assuming the best-fit mass from the \citet{Liu2026} models. The elevated Eddington ratio is in line with the hypothesis of a radiation-driven wind discussed in Sect.~\ref{sec:abs}, and allowing for somewhat larger system masses. The low masses obtained with this model, combined with the stellar mass infered from dynamical arguments for the host galaxy (Sect.~\ref{sec:morph}) set lower limits to the BH-to-stellar mass ratio of $M_{\rm BH} / M_{*} \gtrsim 10^{-4}$--$10^{-2}$, that are compatible with the relations observed in the Local Universe, within the large uncertainties \citep[][]{reines2015}.

\section{Conclusions}\label{sec:conclusions}

In this letter we presented the discovery and spectroscopic confirmation of \panbhs{}, an LRD with an extreme Balmer break at $z=1.731$. The strenght of the Balmer break ($\rm F115W/F814W = 7\pm 1$) is comparable to the most extreme LRDs known, \cliff{} \citep[][]{degraaff20255} and \mombhs{} \citep[][]{naidu2025}. We summarize the observations and our main conclusions as follows.

\begin{enumerate}
    \item We obtained deep VLT/X-Shooter spectroscopy of \panbhs{}. The \Halpha{} emission line is luminous and broad ($L_\Halpha{} = 10^{43}$~\unit{erg.s^{-1}}), and has an unusually strong absorption. \Hbeta{} is detected with $\rm S/N\approx 6$, and we conservatively estimate a lower limit for the Balmer decrement of $\Halpha{} / \Hbeta{} > 9.4$ (at a 3$\sigma$ confidence level), in line with other LRDs in the literature \citep[e.g.,][]{degraaff20251, Nikopoulos2025}.

    \item The absorption trough spans from $-520$~\unit{km.s^{-1}} to $267$~\unit{km.s^{-1}} (at a transmission level of 99\%). We interpret the presence of blue- and red-shifted absorption as produced by a disk wind, analogous to those analysed in the context of broad absorption line quasars or accreting stars. This hypothesis would imply that the source of the optical continuum is likely a thick photospheric disk.

    \item We detect a narrow \Halpha{} component ($\rm FWHM=184\pm 12$~\unit{km.s^{-1}}) which we interpret as probing a host galaxy with $M_* \approx 10^8$~\unit{M_\odot} and $\rm SFR = 2$--3~$M_\odot$. This interpretation is in line with the extended rest-NUV morphology measured in the HST bands ($r_{\rm eff}=1^{+0.5}_{-0.3}$~kpc).

    \item By fitting the synthetic atmosphere models of~\citet{Liu2026}, we estimate a system mass (BH+envelope) of $10^4$--$10^6$~$M_\odot$. The inferred masses, together with the stellar mass inferred from morphology and narrow-emission line dynamics implie BH-to-stellar mass ratios of $10^{-2}$--$10^{-4}$, close to the extrapolated trend in the local Universe \citep[][]{reines2015}.

    \item The confirmation of this source at cosmic noon (magnitude of $\approx 22$ in the $K$ band, H$\alpha$ flux $\approx5\times10^{-16}$ erg s$^{-1}$ cm$^{-2}$) proves the feasibility of detecting extreme LRDs at such epochs with wide-area spectroscopic surveys like Euclid or the forthcoming Nancy Grace Roman Space Telescope.
\end{enumerate}

\begin{acknowledgments}
AT thanks Debasish Dutta and Tamara Bogdanovi\'c for useful conversations about stellar and AGN winds.

We thank the scientific referee for the useful and constructive feedback, which helped improving the quality of this paper.

JM and AT acknowledge funding by the European Union (ERC, AGENTS,  101076224).
The work of CCW is supported by NOIRLab, which is managed by the Association of Universities for Research in Astronomy (AURA) under a cooperative agreement with the National Science Foundation.
APC warmly acknowledges the support of the National Science Foundation through the NSF Graduate Research Fellowship Program. AdG acknowledges support from a Clay Fellowship awarded by the Smithsonian Astrophysical Observatory.

Based on observations made with ESO Telescopes at the Paranal Observatory under programme IDs 116.294D and 116.2AQ0.

This work is based in part on observations made with the NASA/ESA/CSA James Webb Space Telescope. The data were obtained from the Mikulski Archive for Space Telescopes at the Space Telescope Science Institute, which is operated by the Association of Universities for Research in Astronomy, Inc., under NASA contract NAS 5-03127 for JWST. These observations are associated with programs \#2514, \#9433. CCW gratefully acknowledges support for program JWST-GO-2514 provided by NASA through a grant from the Space Telescope Science Institute, which is operated by the Association of Universities for Research in Astronomy, Inc., under NASA contract NAS 5-03127. 
The authors acknowledge the team led by co-PIs R.~Maiolino and F.~D'Eugenio for developing their observing program with a zero-exclusive-access period.

This research is based on observations made with the NASA/ESA Hubble Space Telescope obtained from the Space Telescope Science Institute, which is operated by the Association of Universities for Research in Astronomy, Inc., under NASA contract NAS 5–26555. These observations are associated with program \#15117.

The JWST and HST data presented in this article were obtained from the Mikulski Archive for Space Telescopes (MAST) at the Space Telescope Science Institute. The specific observations analyzed can be accessed via \dataset[doi: 10.17909/ydwx-st06]{https://doi.org/10.17909/ydwx-st06}.

This work is based in part on observations made with the Spitzer Space Telescope, which was operated by the Jet Propulsion Laboratory, California Institute of Technology under a contract with NASA. The Spitzer data used in this work can be found in \dataset[doi: 10.26131/IRSA3]{https://doi.org/10.26131/IRSA3}.

This work was supported by the International Space Science Institute (ISSI) in Bern, through ISSI International Team project \#25-659 `Little Red Dots, Big Open Questions'.

JWST cartoon in Fig.~\ref{fig:wind_cartoon} credit: NASA.

\end{acknowledgments}

\begin{contribution}
AT led the data analysis, and drafted a first version of the manuscript.
JM is the PI of the VLT/X-Shooter programs used, and performed the spectroscopic data reduction.
AW and RPN conducted the photometric search in which the target was selected and AW also performed the morphological analysis.
YM and AT carried out part of the VLT/X-Shooter observations used in this work at the Paranal observatory in December 2025.
AD contributed to the developement of the X-Shooter data reduction pipeline.
All authors contributed to the discussion, interpretation and writing the text.

\end{contribution}

\facilities{VLT (X-Shooter, HAWK-I), JWST (NIRCam, NIRspec), HST (ACS), Spitzer (IRAC, MIPS).}

\software{
    \textsc{Astropy} \citep[][]{astropy:2013, astropy:2018, astropy:2022}, \textsc{NumPy} \citep{numpy2020}, \textsc{SciPy} \citep[][]{2020SciPy-NMeth}, \textsc{pysersic} \citep[][]{pysersic}, \textsc{stpsf} \citep[][]{Perrin2014}, \textsc{lmfit} \citep[][]{lmfit2014}, \textsc{esorex} \citep[][]{esorex}, Claude (used for Python coding; \url{https://claude.ai/}), \textsc{sep} \citep[][]{barbary2016}.
          }

\appendix

\section{Spitzer and VLT photometry}\label{sec:spitzer_phot}

\begin{figure}
    \centering
    \includegraphics[width=0.8\linewidth]{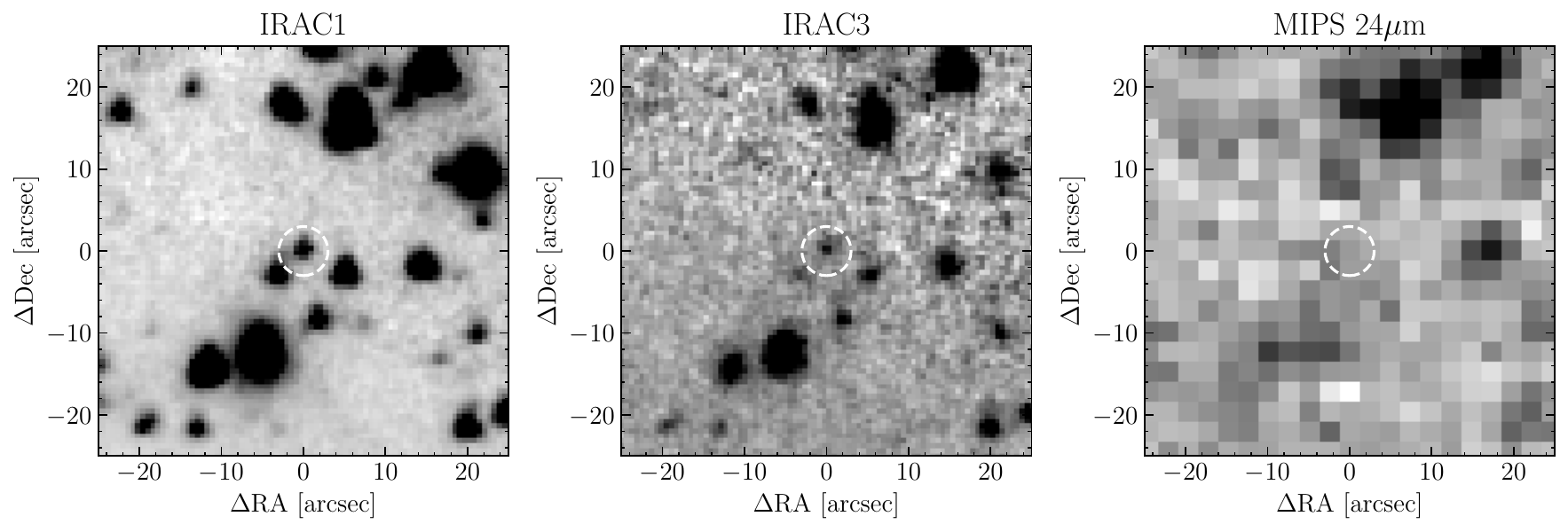}
    \caption{Cutouts of the Spitzer IRAC and MIPS bands with coverage of \panbhs{}. The source is clearly detected in the IRAC bands 1 and 3, and undetected in the MIPS 24$\mu$m channel. For reference, we show a 6\arcsec{} diameter aperture with a white cicle, as used to estimate the 24~\unit{\mu m} upper limit.}
    \label{fig:irac_phot}
\end{figure}

\begin{figure}
    \centering
    \includegraphics[width=0.7\linewidth]{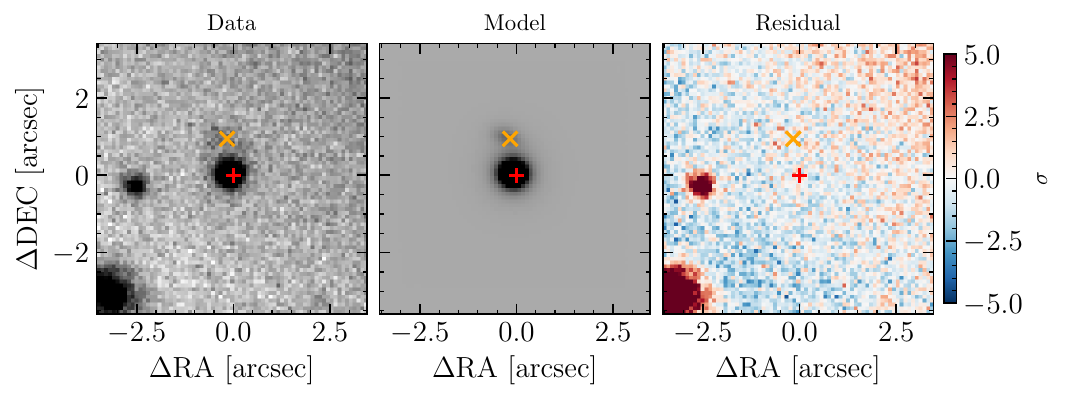}
    \caption{Cutout of of \panbhs{} in the VLT/HAWK-I $K_s$ band. We model \panbhs{} as a point source, and the companion with a Sérsic profile, in order to measure clean fluxes.}
    \label{fig:vlt_img_cutout}
\end{figure}

The data was retrieved from the Infrared Science Archive (IRSA)\footnote{\url{https://irsa.ipac.caltech.edu/frontpage/}}. We downloaded $10\times10$\arcmin{} cutouts, and performed aperture and Kron photometry on Spitzer/IRAC (3.6 and 5.8~\unit{\mu m}) and MIPS (24~\unit{\mu m}) cutouts centred on \panbhs{} using the Python package \textsc{sep} (a Python wrapper around Source Extractor).
We perform background subtraction with the \textsc{SourceExtractor} implementation in Python, \textsc{sep} \citep{Bertin96, barbary2016}.
We measure Kron AB magnitudes of $21.6^{+0.1}_{-0.1}$ and $21.8^{+0.3}_{-0.2}$ for the IRAC 3.6~\unit{\mu m} and 5.7~\unit{\mu m}, respectively. The Kron FWHM were 2.3\arcsec{} and 2.0\arcsec{}, comparable to the typical PSF of IRAC short wavelength channels \citep{Fazio2004}.
\panbhs{} undetected in the MIPS 24~\unit{\mu m} band. Using an aperture diameter of 6\arcsec{} ($\sim 1$ PSF). We obtain a 3$\sigma$ limit magnitude of 19.5.
The flux measurements are broadly consistent with JWST/NIRCam at similar wavelengths (Table~\ref{tab:pan1115_photometry}).

The $50\times50$\arcsec cutouts of the available Spitzer bands are shown in Fig.~\ref{fig:irac_phot}.
Given the large PSF of Spitzer, \panbhs{} is blended with a fainter source roughly one arcsec to the north (see masked region in Fig.~\ref{fig:morph_fit_residuals}). This source corresponts to ID 1114 from the catalog in Weibel et al. (in prep.), with a secure photometric redshift of $z=0.52$. This interloper is much fainter, with $\rm F444W = 25.13$, hence we expect minor contamination to the IRAC fluxes of \panbhs{}. Additional contamination could come from the brigther source a few arcseconds to the south-east that can be spotted in Fig.~\ref{fig:irac_phot}. For this reason, we only use the Spitzer data for qualitative assessment and we do not use it in the model fits presented throughout the paper.

\panbhs{} is also covered by deep archival VLT/HAWK-I imaging \citep[][]{Brammer16}. We downloaded the mosaic from the ESO archive\footnote{\url{https://archive.eso.org/scienceportal/home}}. We estimated the PSF empirically by median stacking 9 bright stars in the field, with a FWHM of 0.28\arcsec{}. \panbhs{} has a faint, extended companion $\approx 1$\arcsec{} to the north, that could affect aperture flux measurements. In order to accurately deblend the \panbhs{} light from the companion, we model \panbhs{} as a point source, and the companion with a Sérsic, all convolved with the empirical PSF. The decomposition is shown in Fig.~\ref{fig:vlt_img_cutout}. We measure a $K_s$-band magnitude of $22.11\pm0.05$ for \panbhs{}, and $25.01^{+0.17}_{-0.19}$ for the companion.

\begin{table}
\centering
\caption{HST, JWST, VLT and Spitzer photometry of \panbhs{}.}
\label{tab:pan1115_photometry}
\begin{tabular}{ccccc}
\hline
Telescope/Instrument & Filter & $\lambda_{\rm obs}$ [$\mu$m] & $\lambda_{\rm rest}$ [$\mu$m]& $m_{\rm AB}$ \\
\hline
HST/ACS & F606W & 0.591 & 0.216 & $27.02^{+0.29}_{-0.40}$ \\
HST/ACS & F814W & 0.831 & 0.304 & $26.31^{+0.15}_{-0.18}$ \\
JWST/NIRCam & F115W & 1.154 & 0.423 & $24.19^{+0.05}_{-0.06}$ \\
JWST/NIRCam & F150W & 1.501 & 0.550 & $22.87^{+0.05}_{-0.06}$ \\
JWST/NIRCam & F200W & 1.988 & 0.728 & $21.87^{+0.05}_{-0.06}$ \\
VLT/HAWK-I & $K_s$ & 2.146 & 0.785 & $22.11^{+0.05}_{-0.05}$ \\
JWST/NIRCam & F277W & 2.776 & 1.017 & $21.79^{+0.05}_{-0.06}$ \\
Spitzer/IRAC & IRAC1 & 3.551 &  1.298 & $21.6^{+0.1}_{-0.1}$ \\
JWST/NIRCam & F356W & 3.565 & 1.306 &$21.75^{+0.05}_{-0.06}$ \\
JWST/NIRCam & F444W & 4.402 & 1.612 & $21.90^{+0.05}_{-0.06}$ \\
Spitzer/IRAC & IRAC3 & 5.730 &  2.095 & $21.8^{+0.3}_{-0.2}$ \\
Spitzer/MIPS & 24\unit{\mu m} & 23.68 &  8.658 & $>\,19.5^\dagger$ \\
\hline
\end{tabular}
\tablecomments{
Rest wavelengths calculated assuming $z=1.731$.
$^\dagger 3\sigma$ limit.
}
\end{table}

\section{HST Morphological fitting}\label{sec:morph_app}

\begin{figure}
    \centering
    \includegraphics[width=0.8\linewidth]{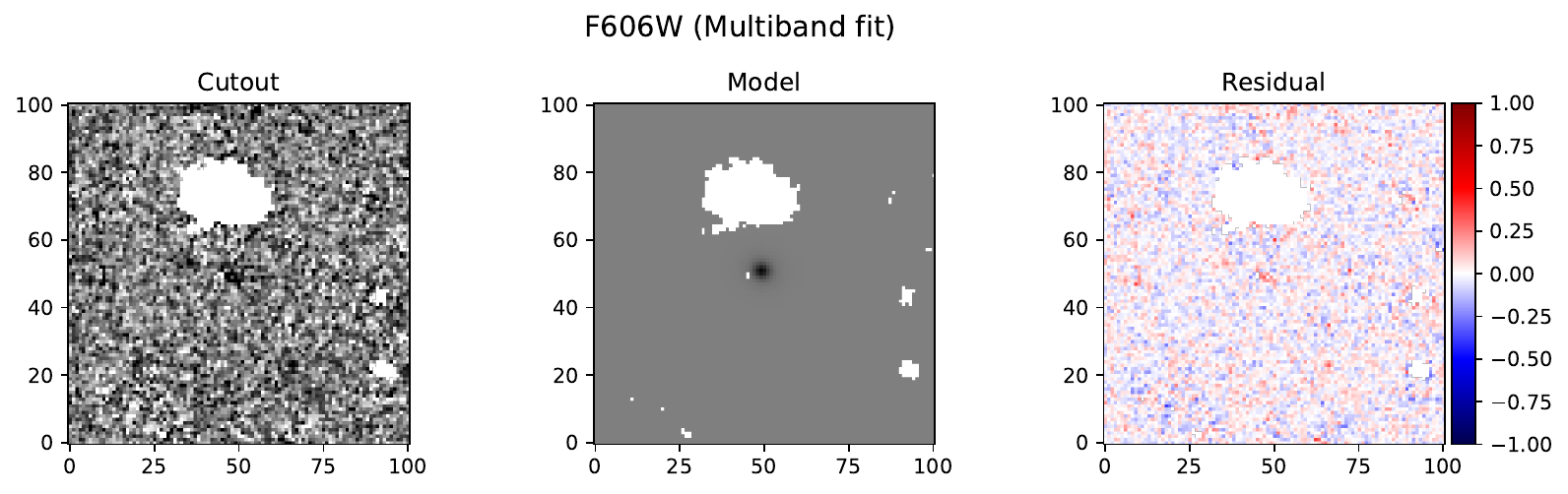}
    \includegraphics[width=0.8\linewidth]{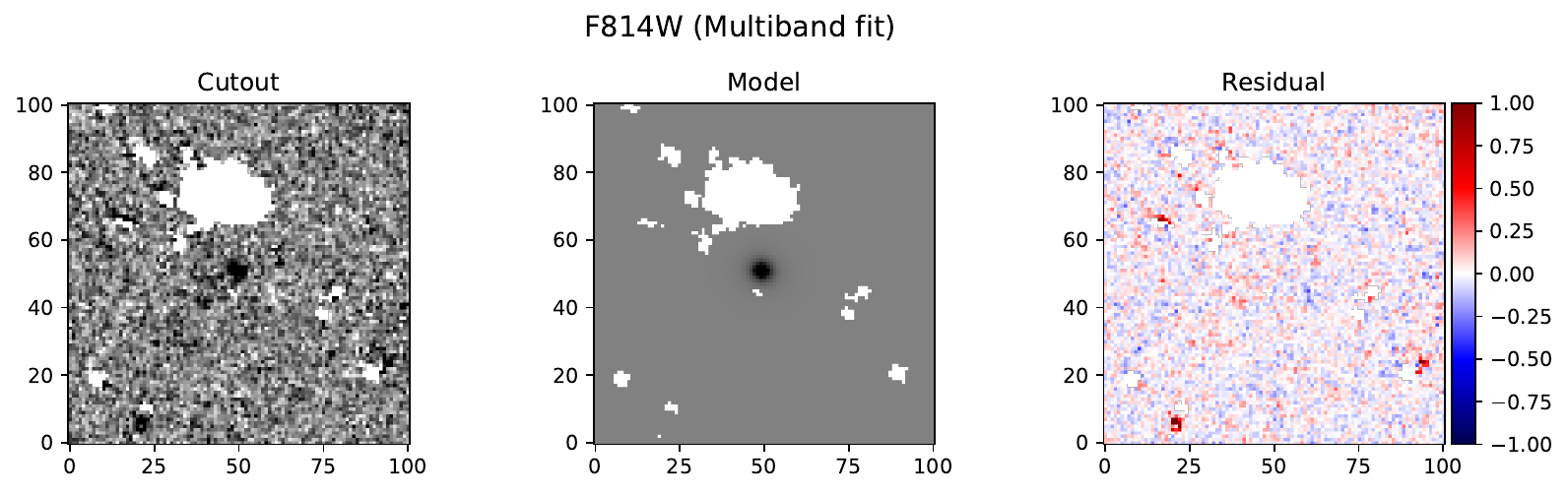}
    \caption{Residuals of the morphological fit in the HST broad bands F606W (top) and F814W (bottom). The all-white pixels are masked out to remove a nearby galaxy, as well as flagged bad pixels.}
    \label{fig:morph_fit_residuals}
\end{figure}

\begin{figure}
    \centering
    \includegraphics[width=1\linewidth]{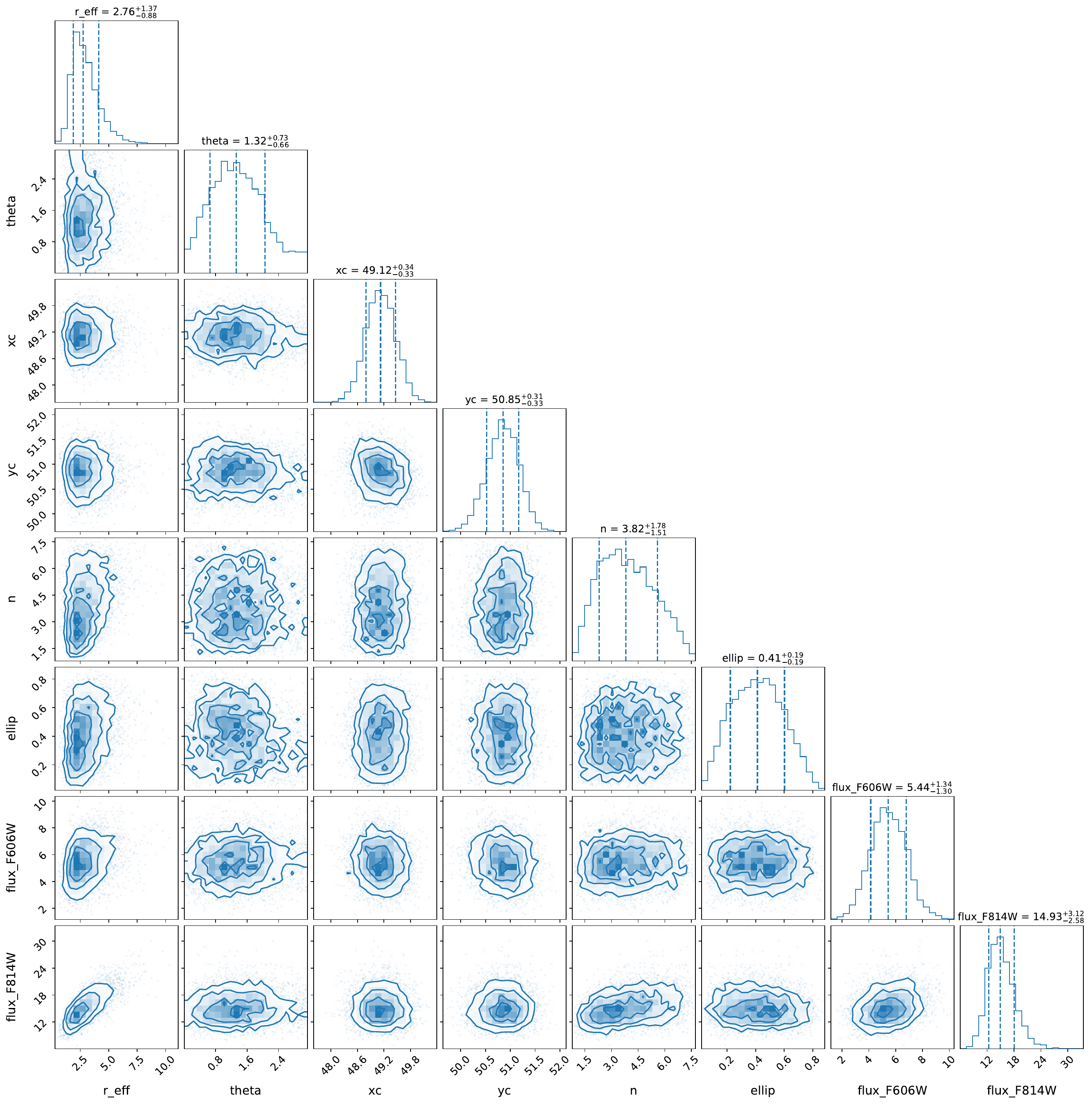}
    \caption{Results from the joint morphological fit in the HST broad bands F606W and F814W.}
    \label{fig:f606w_morph_corner}
\end{figure}

In Fig.~\ref{fig:morph_fit_residuals} we show the best-fit model in the HST F606W and F814W bands, and the residuals of the fit in arbitrary units. In Figs.~\ref{fig:f606w_morph_corner} and~\ref{fig:f606w_morph_corner} we display corner plots with the results of the Sérsic fitting. The source is marginally resolved in both bands, showing consistent sizes and position angles within the uncertainties.

\bibliography{bibliography}{}
\bibliographystyle{aasjournalv7}

\end{document}